\documentclass[reprint,prmaterials,amsmath,amssymb,superscriptaddress,nofootinbib]{revtex4-2}

\usepackage{booktabs}
\usepackage{bigdelim}
\usepackage{graphicx}
\usepackage[utf8]{inputenc}
\usepackage[T1]{fontenc}
\usepackage[english]{babel}
\usepackage{bm}
\usepackage{array}
\usepackage{multirow}
\usepackage{xcolor}
\usepackage{url}
\usepackage{booktabs}
\usepackage{bigdelim}
\usepackage{graphicx}
\usepackage{array}

\newenvironment{hebtab}{%
  \fontsize{8.0}{9.6}\selectfont
  \setlength{\tabcolsep}{1.76pt}%
  \setlength{\arrayrulewidth}{0.4pt}%
}{}

\newcommand{\hebhead}[1]{{\fontsize{5.76}{6.91}\selectfont\textit{#1}}}

\newcommand{\hebrulebold}[1]{%
  \noalign{\global\arrayrulewidth=1.1pt}\cline{#1}%
  \noalign{\global\arrayrulewidth=0.4pt}}
\newcommand{\hebruleouter}[1]{%
  \noalign{\global\arrayrulewidth=0.9pt}\cline{#1}%
  \noalign{\global\arrayrulewidth=0.4pt}}

\newcommand{\hebsection}[2]{\multicolumn{#1}{|l|}{\textbf{#2}}}

\begin{document}

\title{ML-COHP: predicting bond descriptors using latent space of a universal interatomic potentials}

\author{Daniil~S.~Alekseev}
\email{Daniil.Alekseev@skoltech.ru}
\affiliation{Skolkovo Institute of Science and Technology, Bolshoy Boulevard 30, bld. 1, Moscow 121205, Russian Federation}
\author{Ilya~V.~Chepkasov}
\affiliation{Skolkovo Institute of Science and Technology, Bolshoy Boulevard 30, bld. 1, Moscow 121205, Russian Federation}
\affiliation{Bauman Moscow State Technical University, 2nd Baumanskaya Str., 5, Moscow 105005, Russian Federation}
\author{Alexander~G.~Kvashnin}
\affiliation{Skolkovo Institute of Science and Technology, Bolshoy Boulevard 30, bld. 1, Moscow 121205, Russian Federation}

\begin{abstract}
The integrated crystal orbital Hamilton population (ICOHP) is the standard measure of bond strength in a DFT calculation, but every structure needs its own DFT single point and LOBSTER projection, which takes days for a cell of a hundred atoms, such as a grain boundary or a catalytic nanoparticle.
We predict ICOHP from the frozen descriptors of a MACE foundation model.
A symmetric pair head reads the scalar and bond-projected vector channels of the two atoms and corrects a pair baseline built from the element embeddings of the same model.
We tested the model on the high-entropy diboride (Hf,Zr,Ti,Nb,Ta)B$_2$, Pt clusters on defective graphene and Pt/Pd nanoparticles with adsorbed O$_2$.
The error on bonds stronger than 0.5~eV is 1.6--4.1\%, about half that of the Gaussian-process model of Ferreira et al. trained on the same data.
Trained only on high-temperature HEB$_2$ snapshots, the model reproduces the LOBSTER ICOHP of relaxed grain boundary cells to 0.06~eV in about a second, against 5--51 hours of LOBSTER.
The vector channels halve the error compared to the scalar ones, and the baseline keeps the model usable with two training structures.
With geometries from a fine-tuned MACE potential, the model gives the bonding of Pt/Pd particles over a continuous O$_2$ coverage, and DFT on these particles shows that O$_2$ costs a Pt-rich shell several times more metal--metal bonding than a Pd shell.
\end{abstract}

\maketitle

\section{Introduction}

Crystal orbital Hamilton population (COHP) analysis\cite{dronskowski1993,deringer2011} splits the band energy of a solid into contributions of atomic pairs. 
The integration over the COHP up to the Fermi level gives ICOHP, which is used as a descriptor of bond strength.
Using the LOBSTER\cite{maintz2016,nelson2020} package and automated workflows\cite{george2022} it became a routine step after a plane-wave calculation.
The combination of these approaches to mechanics provides an answer to the question of why a solute makes a grain boundary brittle\cite{ito2021,ito2022cusn,ito2022zn}, why a high-entropy diboride is stronger than its binary compounds\cite{lin2026,wang2026heb2}, or how the ductility of a metallic glass or high-entropy alloy alloy depends on its bonding\cite{moitzi2020,borges2024,guo2025}.

But every one of these calculations costs a DFT single point with symmetry turned off on the full $k$-mesh, together with he projection, which is several hours (or days) per cell with a hundred atoms. 
Therefore, COHP is performed on a few relaxed cells, each with a different configuration, while the questions it aims to answer (bonding at defects, surfaces, adsorbate configurations, composition or coverage) require many more structures than LOBSTER can handle.
A machine learning model that can predict ICOHP based on the structure alone could solve this issue.
Ferreira\cite{ferreira2020} trained a Gaussian-process regression on SOAP descriptors\cite{soap} for Zr--Cu--Al metallic glasses and used it on 10\,000-atom cells. 
Machine-learned Hamiltonians\cite{schnorb,phisnet,deeph,hamgnn} are another approach, as COHP can be derived directly from the Hamiltonian.
However, they require training on the full electronic structure of each system, which adds additional costs.
One can use publicly available data for COHP.
Naik et al.\cite{naik2023,naik2026} published a LOBSTER database of 1520 crystals.
MattKeyBond\cite{mattkeybond} gives per-bond ICOHP extracted from Wannier-type orbitals for 36\,000 crystals from the Materials Project\cite{Jain2013MaterialsProject}. Walterbos et al.\cite{walterbos2026} extended the available bonding data to 2542 halide double perovskites, providing spin-resolved ICOHP and ICOBI from HSE06 calculations and LOBSTER projections.
There are three main issues that are missed by considering these data-based approaches.
(1) The model of Ferreira\cite{ferreira2020} uses a single regression for each interaction type within an alloy family, which is tested on random splits of the same cells. It is not clear whether this approach is appropriate for other types of structures.
(2) The authors did not check whether the bulk crystal databases can be used as a prior for clusters or disordered solids.
(3) To our knowledge, none of the ICOHP models have been tested on structures that are absent from their training set, which is a practical limitation since the cells that cannot be afforded in LOBSTER may not be included in the training data.

Here we build the ML-COHP model on top of the frozen node descriptors from a MACE foundation model\cite{mace,macemp0}, trained on millions of DFT configurations of the entire Periodic table.
The model has scalar and vector channels for each atom, and the bond model only needs to learn a readout. 
The bond model is a symmetric multilayer perceptron (MLP) over the descriptors of two atoms, where the vector channels are projected onto the bond.
On top of this, there is a pair baseline that has parameters that are functions of the embeddings of each element in the same model.
This means that the bond model is defined for every possible element pair.

We first measure the accuracy of the model: what it is based on (the baseline, bond projections, foundation model, and pretraining) for amorphous HEB$_2$, the held-out errors of the final model on three systems, and its cost.
We then apply it to multicomponent diborides, where a model trained on amorphous HEB$_2$ snapshots is tested on relaxed bulk and grain-boundary cells calculated with LOBSTER.
We also apply it to Pt clusters on defective graphene support, held out by cluster size, and to Pt/Pd nanoparticles with adsorbed O$_2$ of different saturation ratio, where DFT labels at three coverages show how oxygen affects the metal bonds, and the model with a fine-tuned MACE potential fills in the coverage axis between them.
The main result is that a model trained only on high-temperature HEB$_2$ snapshots reproduces the bonds of relaxed grain boundaries up to 0.06 eV, with an error set by the descriptors, and pretraining does not change this.

\section{Methods}
\label{sec:methods}

% The detailed information about parameters of DFT calculations and LOBSTER, the comparison of foundation models, the training protocols, the fine-tuning of the MACE potential and the cost measurements are given in the Supplementary Materials.

\subsection{DFT calculations}

All density functional calculations are done in VASP\cite{vasp,paw1994,paw1999}. 
Every structure that gives bond labels is recomputed as a single point without symmetry and projected on the pbeVaspFit2015 basis\cite{maintz2016}, either with LOBSTER\cite{maintz2016,nelson2020} or with gpucohp\cite{gpucohp}, which is our GPU implementation of the LOBSTER scheme. 
Test calculations of a 48-atom HEB$_2$ cell with grain boundary are performed by using both codes and give the same ICOHP with an RMSE of $6\times10^{-5}$~eV.

High-entropy boride, which is considered as a study case here stands for (Hf,Zr,Ti,Nb,Ta)B$_2$, HEB$_2$ for short. 
HEB$_2$ cells are computed with PBE\cite{pbe}, energy cutoff is set to 500 eV and no spin polarization is taken into account.
The grain boundary cells are relaxed while the forces acting on the atoms will be less then 0.01 eV/\AA{} at k-points spacing resolution of 0.30~\AA$^{-1}$.
These structures then recomputed at a $\Gamma$-centered mesh with k-point mesh with resolution of 0.16~\AA$^{-1}$ including Gaussian smearing of 0.05 eV for the bulk binary diborides, while the tetrahedron method is used for the HEB$_2$ bulk and all cells with grain boundaries required for projection with LOBSTER 5.1.1. The smearing method for HEB$_2$ and grain boundaries was chosen to shorten the LOBSTER runs; recomputing bulk NbB$2$ and TaB$_2$ with tetrahedra changes their ICOHP by at most 0.013 eV.
The amorphous HEB$_2$ structures are not relaxed using DFT and instead they are calculated using previously trained MACE potential for HEB$_2$\cite{alekseev2026heb2gb}.
For amorphous structures the single point SCF calculations are performed with Gaussian smearing of 0.05 eV and k-point mesh with resolution of 0.30~\AA$^{-1}$.
This information is used for calculations of projections with gpucohp. 
Obviously, the train setup with different k-point spacing and projection software differs from the test setup for HEB$_2$, but the observed metrics of ML-COHP in this and other applications suggest that even if there is some discrepancy in the DFT/COHP labels, it is negligible compared to the accuracy errors of ML-COHP.

The Pt clusters on graphene are computed with PBE and the D3 correction\cite{pbe,d3}, without spin polarization, at the $\Gamma$ point.
The relaxations are performed with energy cutoff of 400 eV and with Methfessel--Paxton smearing of 0.05 eV.
The single points are done with energy cutoff of 300 eV, and Gaussian smearing of 0.2 eV. 

The Pt/Pd particles are computed with RPBE functional and the D3 correction\cite{rpbe,d3}, taking into account spin polarization, at the $\Gamma$ point.
The geometry relaxation is done with energy cutoff of 480 eV and Methfessel--Paxton smearing of 0.05 eV. 
The single point calculations are performed with Gaussian smearing of 0.2 eV and are projected with gpucohp.

\subsection{ICOHP datasets}
\label{sec:labels}

For predicting the relative bond strengths, we gathered the ICOHPs per atom pair integrated up to the Fermi level as descriptors, in eV, with the LOBSTER sign (negative is bonding) and spins summed where a spin-polarization is taken into account. The sources are listed in Table \ref{tab:sources}.

\begin{table*}[!tbp]
\centering
\caption{Sources and DFT parameters of per-bond ICOHP labels. Structures column defines how many structural configurations are in corresponding dataset. Pairs column means how many pairs in said structures with their ICOHP labels are considered. The first three sets, (a), are the applications. The ordered set (b) is used as a test set for HEB$_2$ application. Two public databases (c) are used for research into pretraining capabilities.}
\label{tab:sources}
\begin{hebtab}
\begin{tabular}{@{}|l|r|r|l|l|@{}}
\hebruleouter{1-5}
\textbf{Source} & \multicolumn{1}{c|}{\textbf{Structures}} & \multicolumn{1}{c|}{\textbf{Pairs}} & \textbf{Elements} & \textbf{DFT, COHP software/source} \\
\hebrulebold{1-5}
\hebsection{5}{a.~Applications} \\
\cline{1-5}
Pt$_n$ on defective graphene & 30 & 2\,476 & C, N, Pt & VASP PBE-D3, LOBSTER 5.1.1 \\
\cline{1-5}
Pt/Pd$_{79}$ + $n$O$_2$ & 30 & 19\,526 & O, Pd, Pt & VASP RPBE-D3 with spin polarization, gpucohp\\
\cline{1-5}
amorphous HEB$_2$ & 1\,246 & 443\,936 & B, Hf, Nb, Ta, Ti, Zr & VASP PBE, gpucohp \\
\hebrulebold{1-5}
\hebsection{5}{b.~Test only} \\
\cline{1-5}
ordered bulk and GB cells & 24 & 22\,995 & B, Hf, Nb, Ta, Ti, Zr & VASP PBE, LOBSTER 5.1.1 \\
\hebrulebold{1-5}
\hebsection{5}{c.~Public databases (pretraining)} \\
\cline{1-5}
Naik et al.\cite{naik2023} & 1\,520 & ${\sim}3\cdot10^5$ & 60+ & VASP PBE, LOBSTER \\
\cline{1-5}
MattKeyBond\cite{mattkeybond} & 36\,332 & $3.3\cdot10^7$ & 83 & Quantum ESPRESSO PBE, Wannier-type orbitals \\
\hebruleouter{1-5}
\end{tabular}
\end{hebtab}
\end{table*}

\begin{figure*}[t]
\centering
\includegraphics[width=0.95\textwidth]{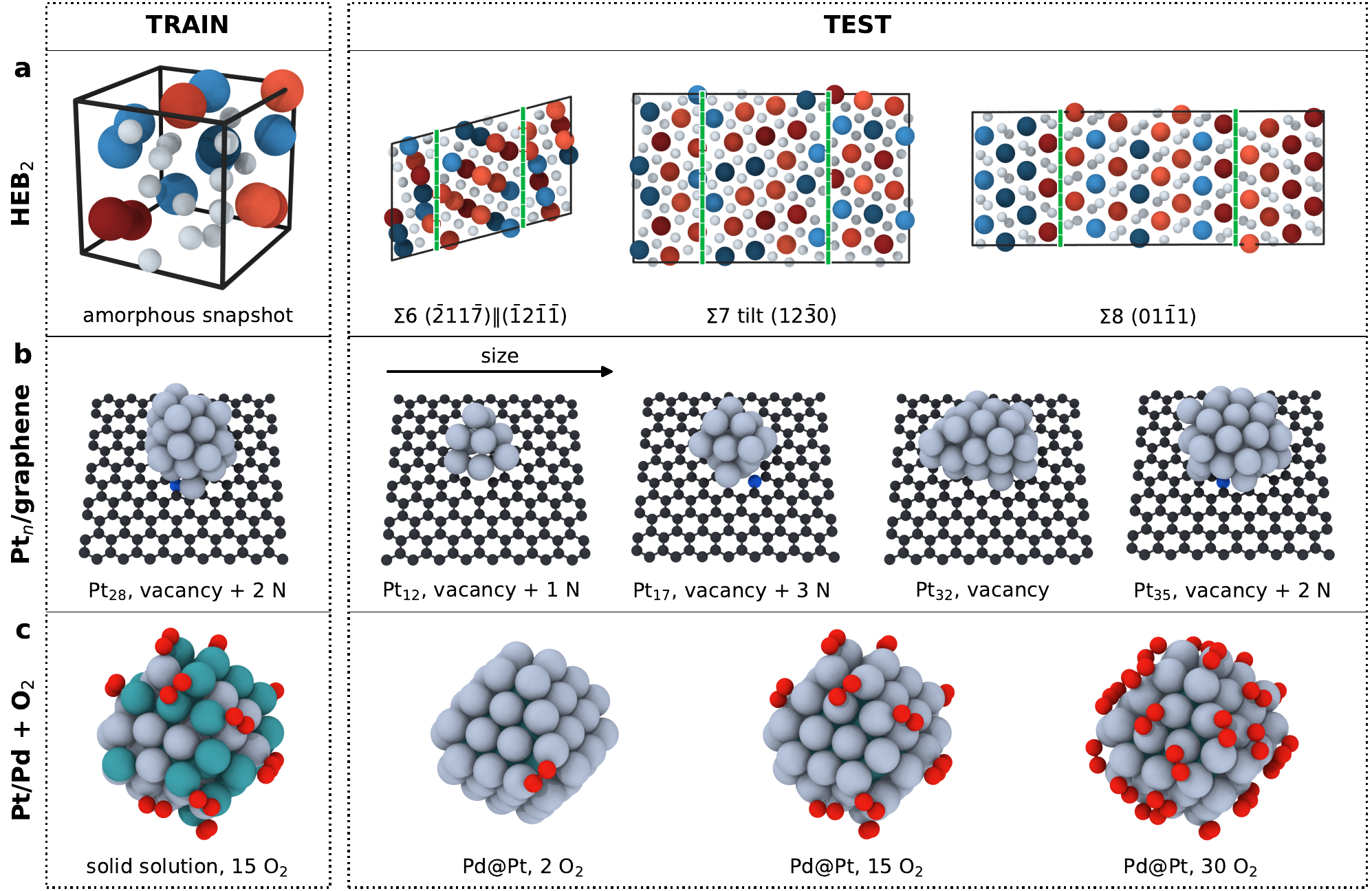}
\caption{Schematic illustration of the studied systems. One structure for training and several test structures per system. 
a) (Hf,Zr,Ti,Nb,Ta)B$_2$, where the model is trained on set of amorphous snapshots (1246 structural configurations) and tested on relaxed ordered cells together with the $\Sigma 6$ asymmetric, $\Sigma 7$ symmetric tilt and $\Sigma 8$ coherent twin boundaries. Structures of considered grain boundaries are shown along their shortest in-plane axis with the boundary normal horizontal. Boundary planes highlighted as dashed green. 
Ti, Zr, Hf are shown in red from light to dark, Nb is dark blue, Ta is light blue, B atoms are small and pale grey.
b) Pt$_n$/graphene, where Pt$_{28}$ on graphene with vacancy and two pyridinic nitrogen atoms is used for training. 
Pt$_{12}$--Pt$_{35}$ on six types of graphene supports, depending on the types of defects, consisting of 30 structures are used for testing held out by the cluster size.
Size of the cluster and types of support are implicitly shown on the figure. Pt atoms are grey, C atoms are black.
c) Pt/Pd + O$_2$ with 30 structures of eleven Pt/Pd$_{79}$ particles with 2, 15 and 30 adsorbed O$_2$ molecules held out by particle type (Pt$_{79}$, Pd$_{79}$, Pt@Pd, Pd@Pt, Pt or Pd core under a shell with 7, 30 or 53 atoms of the other metal, and Pt$_{40}$Pd$_{39}$ solid solution; all shown in Supplementary). 
Pt atoms are shown by light grey, Pd is teal, O is red.
Atomic structures of considered systems are generated using OVITO\cite{ovito}.}
\label{fig:systems}
\end{figure*}

The HEB$_2$ amorphous set contains 1\,246 snapshots, with the number of atoms ranging from 9 to 53 in each cell, which has average dimensions between 4.5 and 9.2~\AA.
Every snapshot is a structure obtained from an MD run at 3000--6000 K with a random composition of two to five metals (nonequimolar compositions are also taken into account) and a new density, simulated with a MACE-MH-1 potential fine-tuned on the PBE data for HEB$_2$\cite{alekseev2026heb2gb}.
The resulting dataset represents a set of amorphous HEB$_2$ configurations, which we refer to below as amorphous snapshots.
The MACE-MH-1 potential only generated the structures with labeling for COHP analysis done in DFT. 
Each snapshot is then recalculated in VASP\cite{vasp,paw1994,paw1999} with PBE\cite{pbe} and projected by using gpucohp\cite{gpucohp}, our GPU implementation of the LOBSTER scheme\cite{maintz2016}with  basis. It is used because it is faster than LOBSTER (Sec.~\ref{sec:cost}) and reproduces its results on a boundary cell. 
The first 508 snapshots are used for the development of the method described in Sec.~\ref{sec:ablation}.

Together with amorphous snapshots the ordered set of structural configurations consists of TiB$_2$, ZrB$_2$, HfB$_2$, NbB$_2$, TaB$_2$, one SQS\cite{sqs} decoration of HEB$_2$ (each has an AlB$_2$-type structure) and three grain boundaries ($\Sigma 6$ asymmetric, $\Sigma 7$ symmetric tilt, $\Sigma 8$ coherent twin with 48--105 atoms) built with pymatgen\cite{pymatgen} and fully relaxed in VASP. 
Single points projections are calculated with LOBSTER 5.1.1 itself.
Boundary atoms are selected as sites within 1.2~\AA{} from the boundary plane.
No structure of this kind is present in any training set, and no model is trained on it.
Instead, it is evaluated after the model had been fixed.

Second case of study is devoted to Pt nanoparticle on graphene support. 
Information about chemical bonding between nanoparticle and graphene depending on type of graphene defects is crucially important for analysis of nanoparticles stability as catalysts for proton exchange membrane fuel cells \cite{banham2017,meier2012,schmies2018,bayan2025}.
Pt clusters of five sizes (Pt$_{12}$, Pt$_{17}$, Pt$_{28}$, Pt$_{32}$, Pt$_{35}$) are considered on six supports: pristine graphene, graphene with one graphitic N, and graphene with a monovacancy with 0--3 pyridinic N, which gives 30 structures of 171--194 atoms.
All considered systems are computed in VASP with PBE+D3 dispersion correction and projected with LOBSTER.
The structures are held by the cluster size: each of the five folds has one size on all six supports (6 structures) used for testing, and the other four sizes (24 structures) are used for training.
Pt$_{28}$ on graphene with monovacancy and two pyridinic-N defects is shown in Fig.~\ref{fig:systems}b as an example of training configuration.

The eleven Pt/Pd particles containing 79 atoms (19 atoms in the core and 60 atoms in the shell) are considered as a third study case.
Different compositions are considered starting from the pure Pd$_{79}$ and Pt$_{79}$, Pt@Pd and Pd@Pt (core@shell nanoparticles).
Here Pt or Pd core has different size with 7, 30 or 53 atoms. 
The last considered type is a Pt$_{40}$Pd$_{39}$ solid solution.
Each structure is relaxed with 2, 15 and 30 adsorbed O$_2$ in VASP with RPBE-D3\cite{rpbe,d3}, with one placement of the molecules per particle and coverage.
The Pt$_{40}$Pd$_{39}$ nanoparticle with 15 adsorbed O$_2$ molecules is used for training of the model. 
The molecules mostly occupy bridge sites, at most one per metal, and stay intact.
Single points of the relaxed structures are projected with gpucohp\cite{gpucohp}, our GPU implementation of the LOBSTER scheme, because it is about 300 times faster than LOBSTER (Sec.~\ref{sec:cost}).

\subsection{Descriptors}

Node features are extracted from a pretrained MACE model with all weights frozen: for each atom, the combined output of both interaction layers is used.
For the medium models this gives 256 scalar descriptors per atom, numbers that do not change when the structure is rotated, and 128 vector descriptors, three-component vectors that rotate with the structure and carry the directions to the neighbours.Eight MACE foundation models were compared\cite{macemp0,salex}. The final model uses OMAT-0 medium\cite{macemp0,omat24}.

\subsection{Pair model}
\label{sec:model}

\begin{figure*}[t]
\centering
\includegraphics[width=\textwidth]{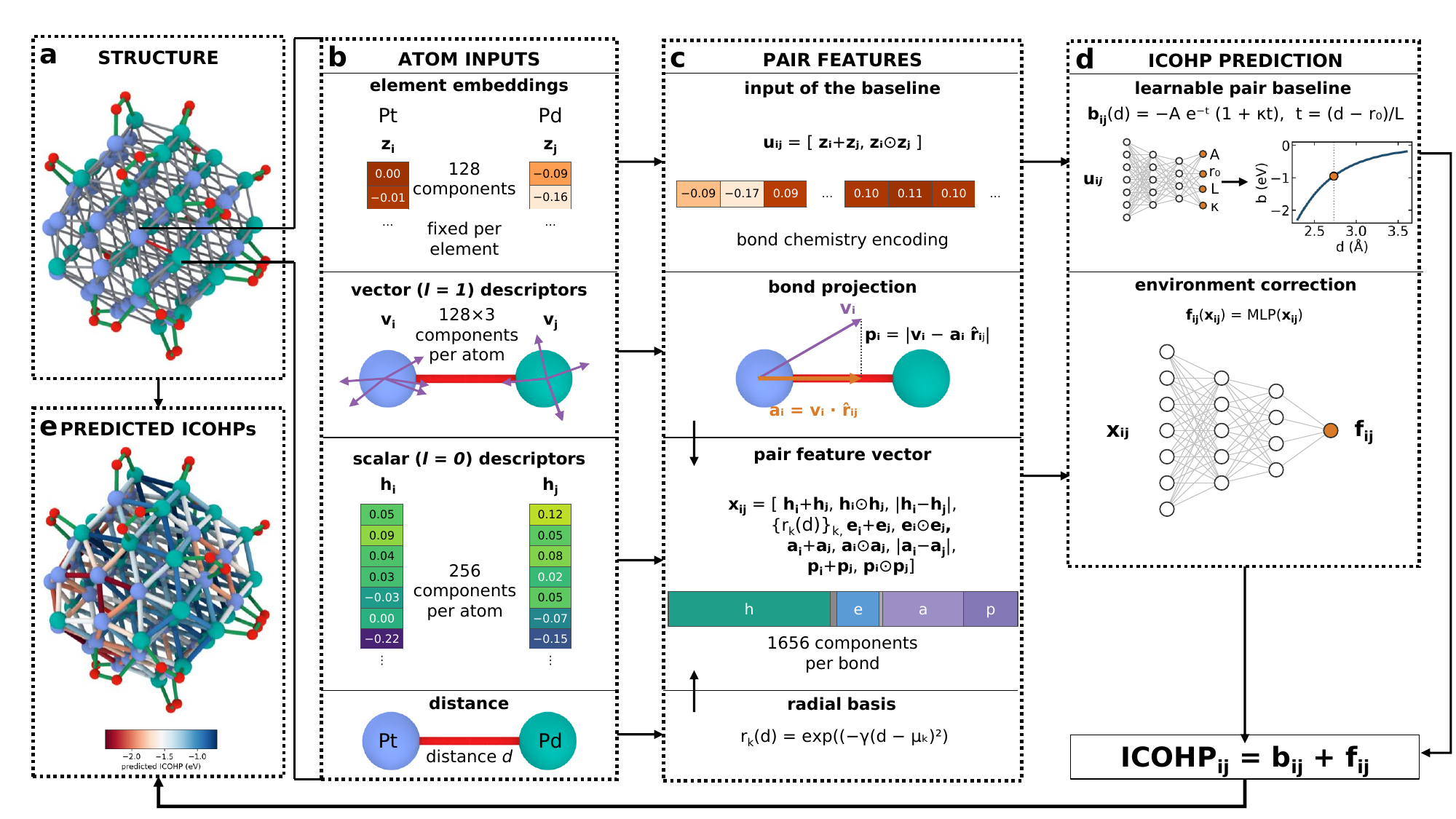}
\caption{Detailed illustration of the pair model for a Pt--Pd bond $i$--$j$ (red) on the surface of Pt$_{40}$Pd$_{39}$ alloy nanoparticle with 15 adsorbed O$_2$.
a) The atomic structure of Pt$_{40}$Pd$_{39}$ alloy nanoparticle. Red is for oxygen, Pt is blue, Pd is green. Considered Pt--Pd bond is highlighted by red. 
b) Atom inputs. The element embeddings $\mathbf{z}$, the vector ($l=1$) and scalar ($l=0$) descriptors of the frozen OMAT-0 model for atoms $i$ and $j$, and the distance $d$ between them. 
c) Pair features. The baseline input $\mathbf{u}_{ij}$, the projections $\mathbf{a}$ of the vector channels on the bond together with the magnitudes $\mathbf{p}$ of their perpendicular parts, the pair feature vector $\mathbf{x}_{ij}$, with its blocks to scale, and the radial basis $r_k(d)$.
d) ICOHP prediction. A small MLP maps $\mathbf{u}_{ij}$ to the four parameters of the baseline function $b_{ij}(d)$, and the MLP $f$ which reads $\mathbf{x}_{ij}$. 
The output scale and shift $\alpha$, $\beta$ are left out of the drawing. 
e) The atomic structure of Pt$_{40}$Pd$_{39}$ alloy nanoparticle where every metal--metal bond of the first shell is colored according to its prediction.}
\label{fig:scheme}
\end{figure*}

For a pair of atoms $(i,j)$ located at a distance $d$ from each other with scalar descriptors $\mathbf{h}$, vector descriptors $\mathbf{V} \in \mathbb{R}^{128\times3}$ and one-hot element codes $\mathbf{e}$ (Fig.~\ref{fig:scheme}) the feature vector is defined as follows:
\begin{align*}
\mathbf{x}_{ij} = \big[\,&\mathbf{h}_i+\mathbf{h}_j,\ \mathbf{h}_i\odot\mathbf{h}_j,\ |\mathbf{h}_i-\mathbf{h}_j|,\ \{r_k(d)\},\ \mathbf{e}_i+\mathbf{e}_j,\\
&\mathbf{e}_i\odot\mathbf{e}_j,\ \mathbf{a}_i+\mathbf{a}_j,\ \mathbf{a}_i\odot\mathbf{a}_j,\ |\mathbf{a}_i-\mathbf{a}_j|,\\
&\mathbf{p}_i+\mathbf{p}_j,\ \mathbf{p}_i\odot\mathbf{p}_j\,\big]
\end{align*},
where $r_k(d)$ defines 32 Gaussians with centers evenly spaced on 0.7--7.0~\AA, $\mathbf{a}_i = \mathbf{V}_i\hat{\mathbf{r}}_{ij}$ and $\mathbf{a}_j = -\mathbf{V}_j\hat{\mathbf{r}}_{ij}$ are the 128 vector channels of each atom projected on the bond toward its partner, $\mathbf{p}_i$, $\mathbf{p}_j$ are the magnitudes of the perpendicular parts.
Every block is symmetric with respect to $i \leftrightarrow j$ and invariant under rotation.
$\mathbf{a}_i$ defines how much of the directional part of the environment of atom $i$ points towards its partner.
This is what distinguishes two bonds with the same length in different orientations, and it is not captured by a scalar descriptor.

The predicted value 
\[
\mathrm{ICOHP}_{ij} = \alpha\,\big[\,b_{ij}(d) + f(\mathbf{x}_{ij})\,\big] + \beta
\],
where $b_{ij}$ is a pair baseline, $f$ is an MLP with hidden layers of 256 and 128 (Sigmoid Linear Unit, SiLU, dropout 0.1, inputs standardized and clipped at $\pm10$ standard deviations), and $\alpha$, $\beta$ are a learned output scale and shift.
Two baselines are compared.
The simpler distance baseline is a cubic polynomial in $d$ per source and element pair, fitted to the training set.
The embedding baseline has a Morse-like form:
\[
b_{ij}(d) = -A\,e^{-t}\,(1+\kappa t), \qquad t = (d - r_0)/L
\],
where the four parameters $(A, r_0, L, \kappa)$ are predicted by a small MLP from $[\mathbf{z}_i+\mathbf{z}_j,\ \mathbf{z}_i\odot\mathbf{z}_j]$ of the frozen element embeddings $\mathbf{z}$ of the same foundation model. 
The baseline is fitted alone for 30 epochs before joint training, and a penalty of $10^{-3}\langle f^2\rangle$ keeps the MLP to correct it.
Since its parameters are functions of embeddings and not entries in a table, the baseline embedding is defined for element pairs that are absent from the training data.

\subsection{Training and evaluation}
\label{sec:protocols}

We train AdamW\cite{adamw} (weight decay $10^{-2}$, batches of 4096 pairs, MSE loss) on a validation split of the training data using early stopping. 
A model is an average of three heads trained from different starting points.
Results are given for several seeds, each seed being a model trained anew from a different starting point.
Training a model on 400 amorphous frames takes about one minute on an RTX 4090 GPU.

The data is split into five folds based on the structure, so none of the training data is used for testing.
For the clusters, the splits are more strict.
Pt/graphene is split based on cluster size, so clusters on six supports are held out together, and Pt/Pd is split based on particle coverages, so all particles with a coverage greater than zero are held out together.
We report the held-out RMSE (mean $\pm$ standard deviation over folds) and the relative error for strong bonds, and the average of $|\Delta|/|\mathrm{ICOHP}|$ over bonds with $|\mathrm{ICOHP}| > 0.5$ eV. 
For the clusters, metal--metal bonds are targets unless specified otherwise, as there are few and varied adsorbate bonds.

Two protocols are compared. 
Scratch, where the model is trained on the application only. 
Pretrained, where the model is first trained on Naik et al.~\cite{naik2023} (pairs up to 4.2~\AA{} with $|\mathrm{ICOHP}| \ge 0.02$ eV).
Here $\alpha$ and $\beta$ are fitted to the application by least squares, and everything is fine-tuned.

\subsection{Maps of the descriptor space}
\label{sec:maps}

For the maps of the applications (Fig.~\ref{fig:umaps}) each bond is described by the geometric part of $\mathbf{x}_{ij}$, which consists of the $\mathbf{h}$, $\mathbf{a}$ and $\mathbf{p}$ blocks.
This means that two bonds are considered to be close if their environments are similar, regardless of their individual elements or length. 
To create the maps, the vectors are standardized and reduced to 50 principal components before being projected using UMAP\cite{umap}. 
Each application has its own unique map, with a minimum distance between points of 0.5 and 200 neighbours.

\section{Results}
\label{sec:results}

\subsection{Accuracy of the model}
\label{sec:accuracy}

\subsubsection{Ablation of the model components}
\label{sec:ablation}

\begin{figure*}[t]
\centering
\includegraphics[width=0.95\textwidth]{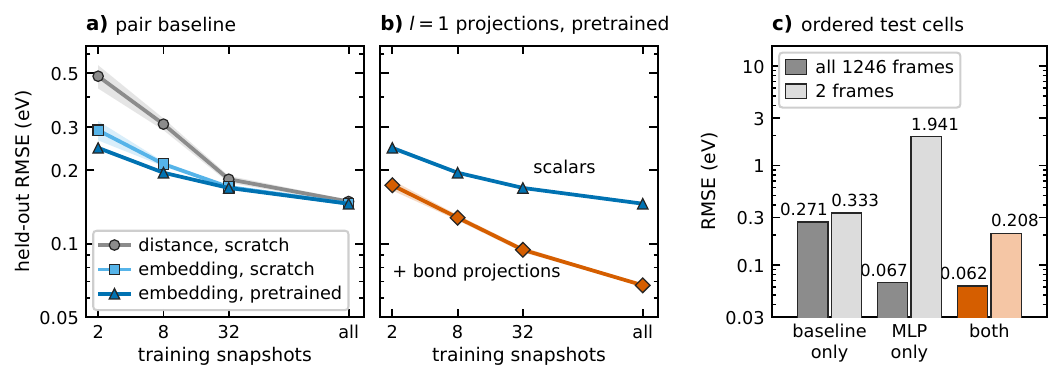}
\caption{Learning curves and ablation on amorphous HEB$_2$. a)~Held-out RMSE depending on the number of training snapshots with scalar descriptors, for the distance baseline trained from scratch and the embedding baseline trained from scratch or pretrained on Naik et al.~\cite{naik2023}.
Five folds by snapshot, colored bands represent the fold standard deviation. 
b)~The same for the pretrained embedding-baseline model without and with the $l=1$ bond projections.\
c)~RMSE on 22\,995 chemical bonds for the ordered HEB$_2$ structures (bulk and grain boundaries) in models trained from scratch using all 1246 amorphous frames (solid bars) or 2 frames (pale bars). 
The embedding baseline, MLP alone and both are shown.
The 2-frame values are mean of three seeds, the full-data ones single runs.}
\label{fig:learning}
\end{figure*}

We built the model on the development set of amorphous HEB$_2$, starting from OMAT-0 scalar descriptors and distance baseline (Fig.~\ref{fig:learning}a). 
The embedding baseline changes the few-shot end of the curve (panel a).
The embedding-baseline model pretrained on Naik et al.~\cite{naik2023}, has half the error of the distance-baseline model at two snapshots, and the pretrained baseline alone, without an MLP, is already better than the entire distance-baseline model (0.34 against 0.49 \, eV).
At full data the two baselines are tied. 
The ablation, shown in Fig.~\ref{fig:learning}c indicates why this matters. 
With all frames the baseline is only worth 8\%, within the change of seed, so in the data-rich regime a plain MLP readout is a valid model. 
With two frames it is not:the MLP alone performs worse at transferring to the ordered cells than predicting the mean, while the baseline alone and the full model give 0.33 and 0.21\,eV.
Therefore, both the element baseline and environment correction play an important role in the final model.

Adding the projected vector channels to the model reduces the error by 53\% at full data (see Fig.~\ref{fig:learning}b), and the improvement increases with the training set.
For strong bonds the relative error drops from 10\% to 4\%. 
Therefore, the vector channels of the foundation model are essential for bonds.

\subsubsection{Held-out errors on the three systems}
\label{sec:final}

\begin{table*}[!tbp]
\centering
\caption{Held-out RMSE on energies in eV of the final model, mean RMSE $\pm$ standard deviation over five folds. Variants (columns) are: \hebhead{scratch, $n=1$} - trained from scratch on one training structure (two for HEB$_2$); \hebhead{pretrained, $n=1$} - pretrained on Naik et al.~\cite{naik2023} and trained on one structure; \hebhead{all} columns use all the corresponding training set instead of one structure. \hebhead{GPR\cite{ferreira2020}, all} is the regression of Ferreira et al.\cite{ferreira2020} trained on the same folds, with at most 10,000 bonds per interaction type.}
\label{tab:final}
\begin{hebtab}
\begin{tabular}{@{}|l|r|r|r|r|r|@{}}
\hebruleouter{1-6}
\textbf{Study case} & \hebhead{scratch, $n=1$} & \hebhead{pretrained, $n=1$} & \hebhead{scratch, all} & \hebhead{pretrained, all} & \hebhead{GPR\cite{ferreira2020}, all} \\
\hebrulebold{1-6}
Pt$_n$/graphene, Pt--Pt & 0.13 $\pm$ 0.02 & 0.17 $\pm$ 0.02 & 0.067 $\pm$ 0.010 & 0.072 $\pm$ 0.013 & 0.112 $\pm$ 0.032 \\
\cline{1-6}
Pt/Pd + O$_2$, M--M & 0.64 $\pm$ 0.60 & 0.14 $\pm$ 0.10 & 0.025 $\pm$ 0.005 & 0.027 $\pm$ 0.006 & 0.056 $\pm$ 0.014 \\
\cline{1-6}
amorphous HEB$_2$, all pairs & 0.47 $\pm$ 0.02 & 0.30 $\pm$ 0.02 & 0.064 $\pm$ 0.001 & 0.063 $\pm$ 0.001 & 0.142 $\pm$ 0.001 \\
\hebruleouter{1-6}
\end{tabular}
\end{hebtab}
\end{table*}

Table \ref{tab:final} gives the errors of the final model for each study case.

For Pt$_n$ on graphene, the folds are by cluster size, so the model is tested on a cluster size absent from training.
With the four training folds the RMSE on Pt--Pt bonds is 0.067~eV, or 2.8\% on strong bonds.
One training structure doubles the error.
Pretraining on Naik et al.\cite{naik2023} does not help here and raises it further, from 0.13 to 0.17~eV.

For the Pt/Pd particles, each fold holds out whole particles with all their O$_2$ coverages.
With the four training folds the RMSE on metal--metal bonds is 0.025~eV (1.6\% on strong bonds).
From scratch, one training particle is enough or not depending on which particle it is (0.64 $\pm$ 0.60~eV), and pretraining lowers this to 0.14~eV.

For amorphous HEB$_2$, the folds are by snapshot.
Each snapshot comes from its own MD run, so a test snapshot shares no trajectory with the training ones, and the RMSE over all pairs is 0.064~eV, or 4.1\% on strong bonds.
With one training snapshot, pretraining lowers the error from 0.47 to 0.30~eV.

At full data pretraining changes nothing beyond the fold spread in any of the three cases.

To compare with the regression of Ferreira\cite{ferreira2020}, we use it SOAP\cite{soap} descriptors with one Gaussian process per interaction type, and with the same kernel parameters and train it on the same folds as our model.
On Ferreira\cite{ferreira2020} Zr--Cu--Al data (16 cells of 100 atoms with 13\,584 pairs, five folds by cell) the regression gives an RMSE of 0.073~eV while  our model gives 0.043~eV.
On our three study cases the error of the regression is 1.7--2.2 times higher than ours (Table~\ref{tab:final}), and on the ordered HEB$_2$ cells it is 0.12 against 0.057~eV.
The regression was limited to 10\,000 training bonds per interaction type, since the cost of a Gaussian process grows as the cube of the training set; with 3000 bonds, the largest set in Ferreira's paper, the error on the ordered cells is 0.20~eV.

\subsubsection{Computational cost}
\label{sec:cost}

\begin{figure}[!tbp]
\centering
\includegraphics[width=\linewidth]{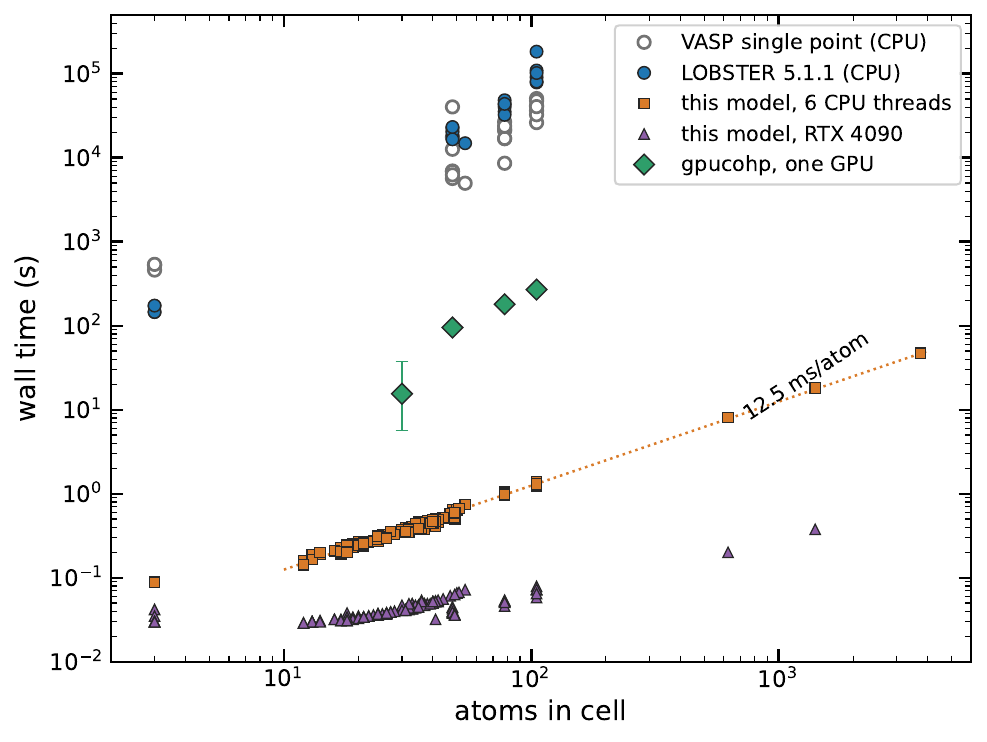}
\caption{Dependence of the wall time per HEB$_2$ structure on its size.
VASP single point calculation with setting for LOBSTER are shown by white circles, LOBSTER v.5.1.1 calculations are shown by blue squares. These simulations are performed for the 24 ordered HEB$_2$ cells (12--20 cores of a shared Intel Xeon). 
The ICOHP model launched on six CPU threads is shown by orange squares, while the same model ran on one RTX 4090 GPU is shown by violet triangles. 
gpucohp calculations based on VASP results are shown by green rhombuses. 
The dotted line is 12.5 ms per atom.}
\label{fig:cost}
\end{figure}

Fig.~\ref{fig:cost} compares the computational cost in terms of the wall time of simulations.
LOBSTER calculations took 5--51 h for the 48--105-atom boundary cells of HEB$_2$, which are on top of 2--14 h of VASP calculations, see blue and white circles in Fig.~\ref{fig:cost}.
Developed model allows one to perform the same type of calculations by taking of about a second on six CPU threads (see orange squares in Fig.~\ref{fig:cost}) and scales linearly at 12.5 ms per atom, 90\% of which is the MACE descriptor pass.
The ML--COHP model allows us to perform the calculations for a 3744-atom cell of HEB$_2$, which takes 47 seconds, while LOBSTER does not attempt to perform such heavy calculations. 
For the structures of HEB$_2$ grain boundaries the model is $3\cdot10^4$ to $10^5$ times faster than LOBSTER alone.
gpucohp is approximately 300 times faster than LOBSTER.
By using gpucohp, the projection stops being a bottleneck, and the model is still 150-200 times faster, and needs no VASP run.
The actual computational cost of the method is the training set, which required 184 h of VASP calculations for the 1246 amorphous HEB$_2$ frames and 5.5 h of gpucohp.
It should be noted that these costs for training have to be paid for each material family, but this still leaves the possibility of performing COHP analysis on MD-scale configurations and trajectories.

\subsection{Applications}
\label{sec:applications}

\subsubsection{High-entropy ceramics and their defects}
\label{sec:unseen}

Transition-metal diborides MeB$_2$ melt above 3000~K. 
They are hard and resistant to oxidation, making them suitable for the leading edges of hypersonic vehicles and rocket nozzles\cite{fahrenholtz2007refractory}. 
With Hf, Zr, Ti, Nb and Ta mixed on one metal sublattice, high-entropy diboride HEB$_2$ is single-phase and about as hard as the binary average, and oxidizes more slowly than all of them except but HfB$_2$\cite{gild2016}.
As a sintered ceramic it is not a single crystal, so the strength and hardness of diborides depend on grain size\cite{fahrenholtz2007refractory,yang2024effect}, and the grains of HEB$_2$ grow five times slower than those of ZrB$_2$, which was attributed to sluggish diffusion and a boundary energy reduced by segregation\cite{cheng2023grain}.
Which metals a boundary takes may be a question of bonding, since at a boundary the bonds are reformed with other lengths and angles.
COHP analysis has already supported the bonding origin of solute segregation to grain boundaries in Fe\cite{ito2021}. 
But every such analysis is a DFT single point and a projection per cell.
So it covers a handful of boundaries with one decoration of the metal sublattice each.
Segregation needs larger cells and many decorations, and a model that predicts ICOHP from the structure alone gives the bonds of each cell from one MACE evaluation.
Here we train ML--COHP model only on amorphous snapshots and tested on relaxed boundary cells.

\begin{figure*}[t]
\centering
\includegraphics[width=0.8\textwidth]{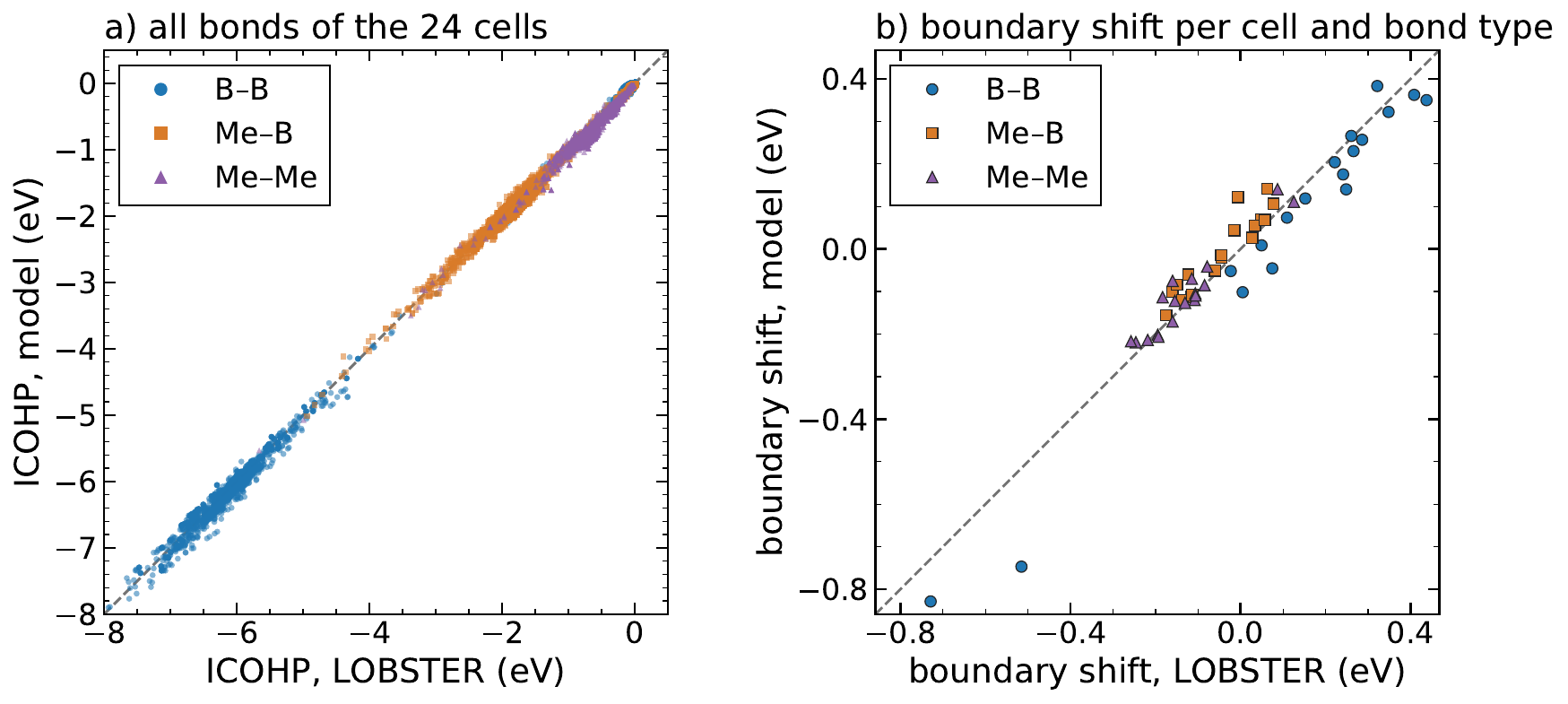}
\caption{The model trained only on the 1246 amorphous HEB$_2$ snapshots, tested on the 24 relaxed bulk and grain-boundary cells computed with LOBSTER v.5.1.1 (mean prediction of three seeds). 
a)~Predicted ICOHP against LOBSTER for all 22\,995 pairs within a cutoff of 4.2~\AA colored by bond type. RMSE is 0.057~eV. 
b)~Boundary shift: for each of the 18 boundary cells and each first-shell bond type, the mean ICOHP of bonds with at least one boundary atom minus the mean of bulk--bulk bonds in the same cell.
The dashed lines are $y = x$.}
\label{fig:gb}
\end{figure*}

The hardest test is the ordered set of relaxed cells at 0 K with single-metal diborides among them projected by using LOBSTER v.5.1.1. 
The model which is trained on all amorphous frames gives an RMSE of 0.057~eV over all 22\,995 pairs (Fig.~\ref{fig:gb}a.
The deviation between seeds is 0.054--0.063~eV for each seed alone.
A low global error does not necessarily mean that the boundary is reproduced correctly, so we compared the mean ICOHP of the bonds that touch the boundary with that of the bulk bonds for each cell. for every cell type and bond type. 
These boundary shifts range from $-0.7$ to $+0.4$ eV in LOBSTER, and the model reproduces these shifts with a correlation of 0.96 and an RMS error of 0.06 eV (see Fig.~\ref{fig:gb}b).

\begin{figure*}[t]
\centering
\includegraphics[width=\textwidth]{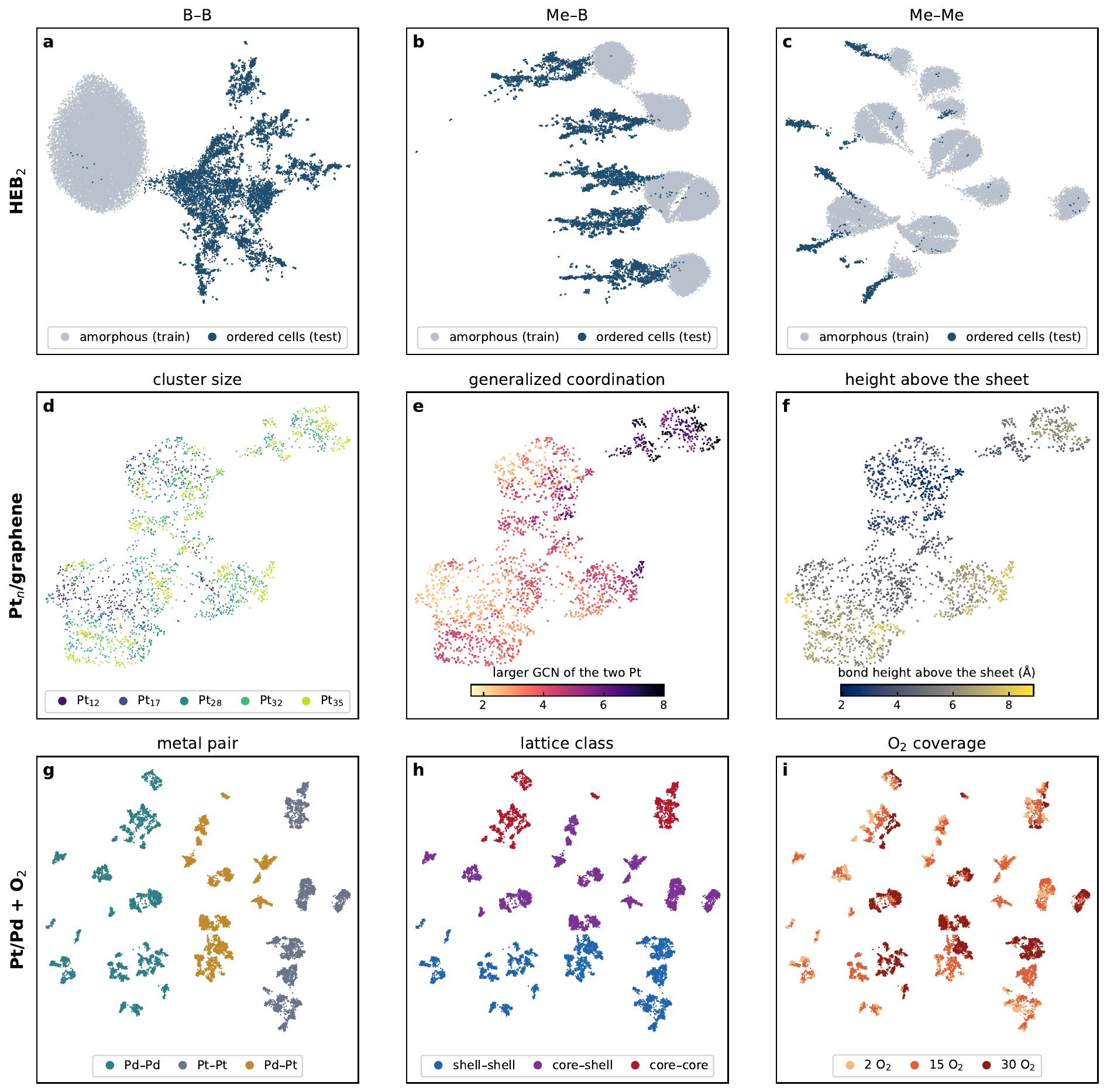}
\caption{UMAP visualization of the geometric bond vectors for each study case (200 neighbors). 
~HEB$_2$ case with visualized a) B--B, b) metal--boron and c) metal--metal bonds. 
20\,000 random pairs (out of 443\,936) of the amorphous HEB$_2$ training configurations (gray) and every pair of the 24 ordered test cells (dark blue; 22\,995 such bonds) are analyzed . 
Pt$_n$/graphene case with all Pt--Pt pairs of Pt$_n$ on graphene, colored by d) the cluster size, e) the larger generalized coordination number of the two Pt atoms (neighbors are only considered) and f) the distance from the bond midpoint to the graphene sheet. 
Pt/Pd + O$_2$ case with g) all metal--metal pairs within 4.2~\AA{} of the 30 DFT Pt/Pd structures, colored by the metal pair, h) the lattice class (core or shell sites of the two metals) and i) the O$_2$ coverage.}
\label{fig:umaps}
\end{figure*}

To analyze the configurational space covered by the training set and the regions that should be predicted by the model, we use the UMAP visualization of the geometric bond vectors for each study case (200 neighbors).
Within HEB$_2$ (Fig.~\ref{fig:umaps}a--c) the bonds of ordered cells lie outside the amorphous set, see gray regions w.r.t. blue one.
Metal--boron and metal--metal bonds  extend from the edge of amorphous bonds in the same pair, while boron--boron bonds form a region next to the amorphous cloud. 
Amorphous frames contain same metal--boron pairs at the same distance, so pair curves are learned.
What the ordered cells change is  coordination, which is handled by environment term in the model.

High-temperature snapshots of structures with a few tens of atoms sample the local environment of relaxed grain boundaries well enough that a boundary-specific DFT is not needed to train the bond model.
This training set must cover all possible bond types over a relevant range of bond lengths and chemical environments.

\subsubsection{Pt nanoparticles on a defective graphene support}
\label{sec:support}

Pt nanoparticles on carbon support are the commercially used catalysts at electrodes of proton exchange membrane fuel cells\cite{banham2017}, and at cathodes.
Both of them requires Pt because of the high overpotential of oxygen reduction\cite{norskov2004,stephens2012}. 
During operation the catalyst loses active surface area as the particles may dissolve, its surface may degrade and redeposit on larger ones, coalesce or detach from the support\cite{meier2012,sandbeck2020}. 
Doping the carbon with nitrogen makes the catalyst more durable\cite{schmies2018,bayan2025}, and DFT attributes this to a stronger adhesion of Pt at pyridinic-N defects\cite{bayan2025}.
But adhesion is one number per particle, while the defect also changes the Pt--Pt bonds inside the nanoparticle.
In order to analyze these internal changes the  COHP analysis can be used for each cluster on each support, i.e. a single point DFT calculation and an LOBSTER projection for each one. 
The model allows one to obtain information about the chemical bonding of nanoparticles of bigger sizes then those present in the training set without running DFT and LOBSTER.

For the Pt clusters on graphene, the folds are determined by the cluster size. 
Therefore, each test cluster has a size that the model has never seen before, on all six supports simultaneously.
With the other four sizes in training procedure the error in the prediction of ICOHP of Pt--Pt bonds is 0.07 eV, 3\% on strong bonds (Table \ref{tab:final}). 
With one labeled structure it is about twice as much, and pretraining according to Naik et al.~\cite{naik2023} makes it worse rather than better.

The UMAP visualization of Pt--Pt bonds are shown in Fig.~\ref{fig:umaps}d--f, which is ordered by the height of the bond above the sheet and the coordination of its atoms, with much less emphasis on the cluster size. 
A held-out size, therefore, shares its bond environment with the sizes in training. 
That is why it is predicted as well as they are.

\subsubsection{Saturation of bimetallic nanoparticles with adsorbates}
\label{sec:ptpd}

Pt and Pd nanoparticles serve as catalysts for hydrogenation and other chemical processes and for the cathodes of proton-exchange membrane fuel cells\cite{suo2022,zhao2022,chen2015}, and Pt--Pd catalysts with the activity of Pt would save part of the scarce Pd\cite{guterman2024,guterman2025}. 
The activity and durability of such catalysts depend on the composition, structure and size of the particles\cite{suo2022,palaniyandy2025}, and the same two metals give a solid solution or a core--shell particle depending on whether their precursors are reduced together or one after the other\cite{antolini2003,lin2025,li2023}.
Alloying acts on the catalysis through the metal.
Alloying changes the lattice parameter and the electronic \textit{d} states of Pt and changes the strength of O and OH adsorption\cite{stephens2012,lim2023}.
Here we look at the opposite direction, i.e. how the metal--metal bonds of the shell, core and interface between them change upon the saturation of the surface with adsorbed O$_2$ and whether this depends on which metal is on the outside. 
It is a per-bond question, and by using LOBSTER each coverage ratio or each position of the molecule on the surface requires additional DFT calculation.
Our developed model provides information about bond strengths based on the structure alone.
This means that we can follow one particle over the entire coverage range with multiple positions of molecules per coverage area.

Each metal atom in the Pt/Pd particle is positioned at a specific site in the reference lattice and remains there at all coverages.
A bond between two neighboring sites is therefore core--core, core--shell or shell--shell, and a particle has 336 such nearest-neighbor bonds, which we follow from 2 to 30 O$_2$ regardless of their current length.
A change with coverage is always taken within one particle, between the same lattice bonds.
We analyze only the metal--metal bonds; the model predicts them with an error of 0.025~eV (Table~\ref{tab:final}).

Fig.~\ref{fig:umaps}g--i shows how the model sees these bonds.
On the map of all metal--metal pairs of the 30 DFT structures every island holds one metal pair (Fig.~\ref{fig:umaps}g) and one lattice class (Fig.~\ref{fig:umaps}h).
An island mixes bonds of several particles, so a bond is placed by its local environment and not by the particle it comes from.
The O$_2$ coverage separates the bonds only partly (Fig.~\ref{fig:umaps}i): inside an island the bonds at 30 O$_2$ often gather on one side.
The lattice class is thus the natural unit for comparing particles, and below ICOHP is averaged over the bonds of one class.

\begin{figure*}[!tbp]
\centering
\includegraphics[width=0.95\textwidth]{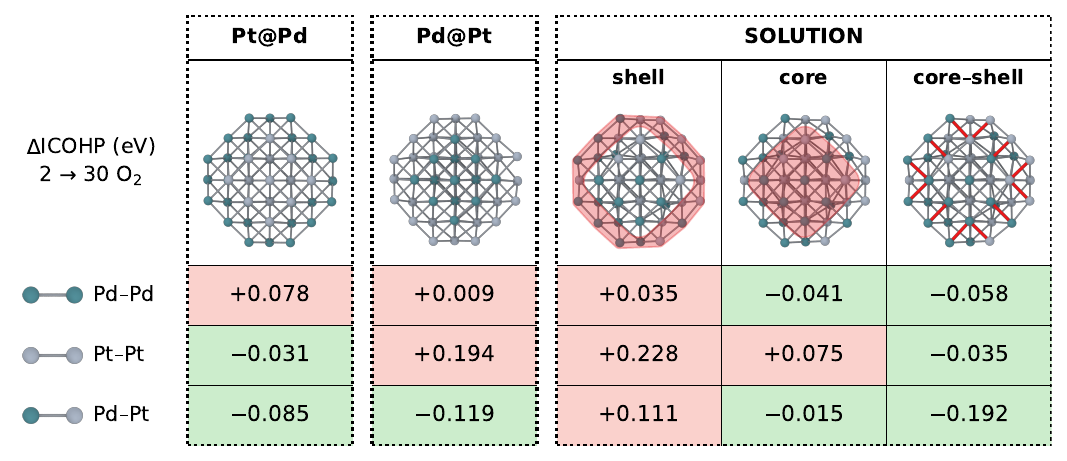}
\caption{Change of the ICOHP (eV) values for M--M bond in Pt/Pd nanoparticle with adsorbed oxygen from 2 to 30 O$_2$ in three particles averaged over the same lattice bonds. 
Columns: Pt@Pd (Pt core, Pd shell), Pd@Pt (Pd core, Pt shell), and the Pt$_{40}$Pd$_{39}$ solid solution restricted to the shell--shell bonds (shell shaded in red), the core--core bonds (core shaded in red) and the core--shell bonds (red sticks).
Headers: OVITO renders of the structures with 2 O$_2$ adsorbed; they are rendered without O, cut through the central atom by a \{100\} plane; Pt is grey, Pd is teal. Green cells show negative values (stronger bonds), red cells show positive values (weaker bonds).}
\label{fig:ptpdtables}
\end{figure*}

First we analyzed the DFT results themselves. Results are present on Fig.~\ref{fig:ptpdtables}. 
We considered three particles of the same size with different arrangements of the metals, Pt@Pd, Pd@Pt and the Pt$_{40}$Pd$_{39}$ solid solution, and the change of ICOHP between 2 and 30 O$_2$ on the surface for every metal pair and lattice class.
In all three particles the shell bonds weaken and the core--shell bonds strengthen.
How much the shell weakens depends on the metal: a Pt--Pt bond in the shell of Pd@Pt or of the solid solution loses about 0.2~eV, a Pd--Pd bond in the shell of Pt@Pd 0.08~eV.
By its metal--metal bonds, the particle with a Pd shell is therefore the most stable under O$_2$ and the one with a Pt shell the least: O$_2$ costs a Pt--Pt shell bond 2--3 times more than a Pd--Pd one, while the interface bonds strengthen in both.
On PtPd alloy surfaces, DFT likewise finds that adsorbed oxygen draws Pd to the surface\cite{dianat2008}.

\begin{figure*}[!tbp]
\centering
\includegraphics[width=0.95\textwidth]{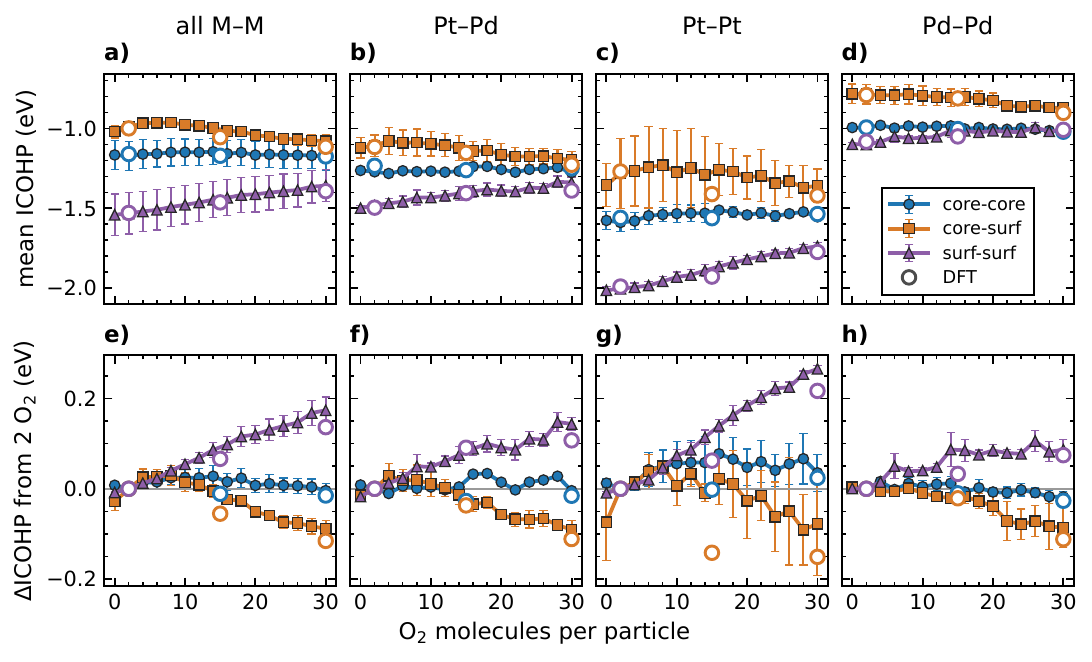}
\caption{Coverage scan performed for structures relaxed with the fine-tuned MACE potential, ICOHP of the lattice bonds is calculated from the held-out model (each nanoparticle predicted by the fold model that did not see it), against the number of O$_2$ adsorbed on the surface of the particle, for the eight particles calculated with DFT at all three coverages.
The scan is performed for all M--M bonds, Pt--Pd, Pt--Pt and Pd--Pd (graphs are ordered is columns). 
a--d)~Mean ICOHP of core--core, core--shell and shell--shell bonds, averaged over the bonds of a structure, the four placements of O$_2$ on a particle and the eight particles, with the standard error over particles.
e--h)~Change of the particle mean from 2 O$_2$. 
Open circles show DFT results for the same eight particles.}
\label{fig:scan}
\end{figure*}

Three particles are a small sample, so Fig.~\ref{fig:scan} repeats the comparison for the eight particles with DFT at 2, 15 and 30 O$_2$ (open circles).
The trend is the same: between 2 and 30 O$_2$ the shell--shell bonds weaken by 0.14~eV and the core--shell bonds strengthen by 0.12~eV, in all eight particles, while the core--core bonds stay where they were (Fig.~\ref{fig:scan}e).
The shell atoms, once bonded to oxygen, presumably move part of their metal bonding from each other to the core, as bond-order conservation would suggest.
Pt--Pt shell bonds lose the most, Pd--Pd the least (Fig.~\ref{fig:scan}f--h). The only discrepancy to DFT results, in Pt--Pt bonds for core-surface class (Fig.~\ref{fig:scan}g), we explain by the MACE optimized geometry which systematically differs to DFT one in this case. This disagreement could be solved with more training data for potential, but for this current application and demonstration we deem the results accurate enough.

\section{Discussion}

The closest previous model, proposed by Ferreira\cite{ferreira2020}, is a Gaussian-process regression per interaction type tested on random splits of its training cells.
Compared to our architecture, it has 1.7--2.2 times higher error on all four data sets (Section~\ref{sec:final}).
We proposed a single model for all 21 type of atomic pairs of six elements and tested on a class of structures absent from training.
We tested the pretraining with respect to Naik et al.\cite{naik2023,naik2026} approach who use LOBSTER output as features, where LOBSTER can be run.
We did not compare our work with other local descriptors (SOAP or ACE with the same head) instead of MACE scalar ones, and machine learning Hamiltonians\cite{schnorb,phisnet,deeph,hamgnn} would give COHP directly, but at a higher training cost.

The accuracy is determined by the descriptors.
The bond projections and the choice of the foundation model contributed most to the improvement, and the next step may come from the $l = 2$ channels, which none of the eight models we compared have.
Of the MACE models released for materials, the MP-0b2 large and one MP-0 variants carry them ($128\times0e + 128\times1o + 128\times2e$). Another possibility is to take descriptors that have explicit bond features, such as ALIGNN\cite{alignn}, Allegro\cite{allegro} or PET\cite{pet}. The information carried by any chosen descriptor could be analyzed by feature reconstruction errors\cite{chorna2026latent}.
Pretraining on bulk-crystal databases is only worth trying when there are one or two labeled structures, and even then, it did not help with a specific Pt/graphene case. 
Another possibility is that there may not be enough data available.
For example,  according to ICOHP predictions there is no Pt--Pt pair of Naik et al.\cite{naik2023} which is stronger than 0.2~eV, while our Pt--Pt bonds reach 2.8~eV.
The optimal way to gather the data for the possible database might be to use farthest point sampling in the space of bond vectors, as Chorna et al.\cite{chorna2026cosmo} did in the latent space of universal potentials.
For the Pt/Pd case, two things are needed in addition to the ICOHP model: a potential to maintain the short-range repulsion from the foundation model and the same type of bonds in each structure.

The amorphous labels may have a random error of up to 0.04 eV per bond due to the coarse $k$-mesh, and the DFT parameters on train and test datasets differed.
Therefore, the held-out errors represent upper bounds of the model error.
The DFT analysis of the particles includes one O$_2$ placement per particle and the coverage is uniform.
No two particles have the same composition.
Transfer to another structure class within the same chemistry was tested, and it was found that the bulk prior fails.
Therefore, the current model needs to be trained for each material family separately.
Transferring to another geometry from amorphous to ordered defects, works in our model.

\section{Conclusions}

In conclusion, model with our proposed architecture, a symmetric pair of head-on frozen MACE descriptors, predicts per-bond ICOHP with an error of 1.6--4.1\% on bonds with ICOHP values higher than 0.5 eV (strong bonds) for Pt clusters on graphene, Pt/Pd particles with O$_2$ and amorphous HEB$_2$, held out by the cluster, particle and type of structural configurations. 
The vector $l=1$ channels projected on the bond halves the error, and the baseline pair from the element embeddings keep the model usable with pretraining and one training structure. The effects of pretraining disappear when there are many application-specific training structures.
Trained only on high-temperature structural configurations of HEB$_2$, the model reproduces the bonds of relaxed diboride grain boundaries to 0.06 eV, including their shift at the boundary.
For Pt/Pd particles, DFT shows that O$_2$ weakens the shell bonds, strengthens the core--shell bonds and costs a Pt-rich shell several times more metal bonding than a Pd shell.
The model with a fine-tuned potential reproduces these trends over a continuous coverage.
The model takes about a second per hundred-atom cell to run on a CPU, instead of days using VASP and LOBSTER.
This makes it possible to perform computational analysis of chemical bonding along molecular dynamics or Monte Carlo simulations, on the trajectories of thousands of atoms at an affordable cost.

\section*{Data and code availability}

The trained models and train/test datasets will be made available at \url{https://doi.org/10.5281/zenodo.23770255}. gpucohp is at \url{https://github.com/daneKKK/gpucohp}. MattKeyBond is at ScienceDB, doi:10.57760/sciencedb.38081, and the data of Naik et al. at Zenodo records 8091844 and 8092187.

\section*{AI-usage}

During the preparation of this work the authors used Claude Opus (Anthropic) to edit the text of the manuscript: keep the track of presented numbers and metrics and help in describing the computational experiments done in the work. All results were rechecked by the authors, who take full responsibility for the content of the publication.

\bibliography{refs}

\end{document}

\title{Supplementary Information for ``ML-COHP: predicting bond descriptors using latent space of a universal interatomic potential''}
\author{D.~S.~Alekseev}
\author{I.~V.~Chepkasov}
\author{A.~G.~Kvashnin}
\affiliation{Skolkovo Institute of Science and Technology, Moscow, Russia}
\maketitle

Section, figure and table numbers without the prefix S refer to the main text.

\section*{Note S1. The two public databases}

The two public databases share 1286 Materials Project ids, 181 of them with identical cells, i.e. 21\,076 bonds computed both ways. For the 5976 of them with $|\mathrm{ICOHP}_\mathrm{LOBSTER}| > 0.1$ eV a linear fit gives:
\[
\mathrm{ICOHP}_\mathrm{LOBSTER} \approx -0.18 + 2.13\,\mathrm{ICOHP}_\mathrm{MKB}
\]
, where MKB stands for MattKeyBond. The correlation is $r = 0.94$ but the residual is 0.67 eV, beyond 3.5~\AA{} the correlation is gone ($r = 0.04$), and per-pair offsets reach $-6.5$ eV for B--O and $-2.9$ eV for Al--O. So the two databases are not on one scale, and we never merged them by rescaling; each entered pretraining as its own source.

\section*{Note S2. Foundation models}

\begin{figure*}[!htbp]
\centering
\includegraphics[width=0.95\textwidth]{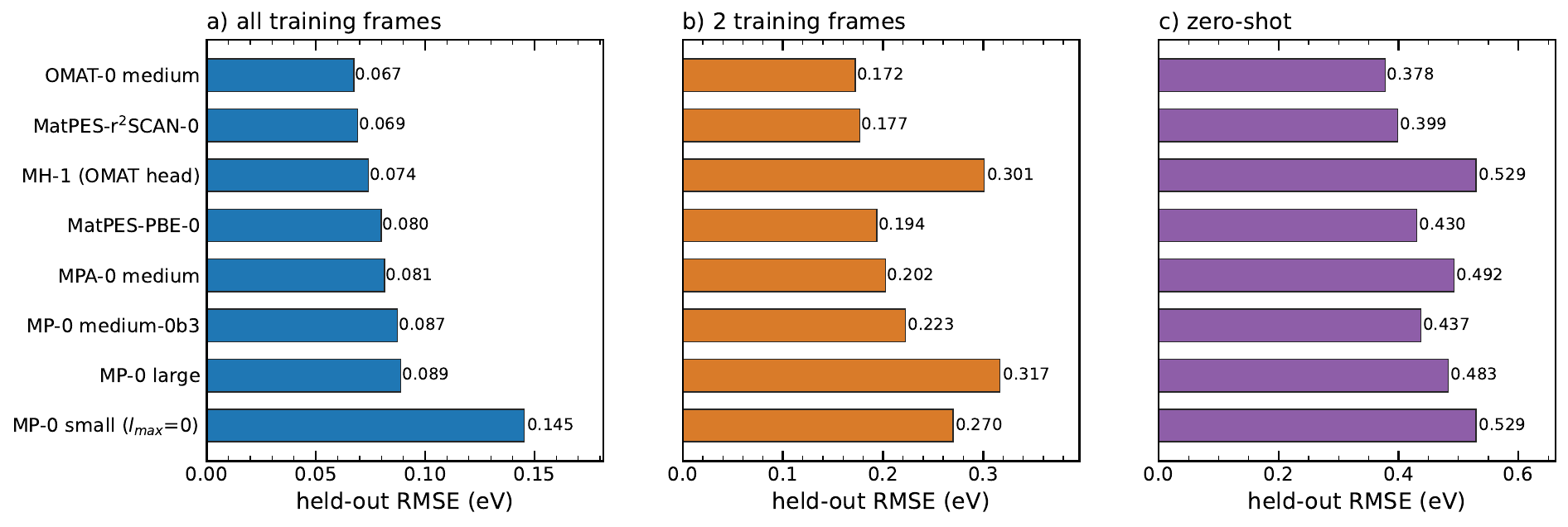}
\caption{Foundation models compared as descriptor sources for the same pair model (embedding baseline, Naik pretraining, bond projections where $l=1$ channels exist) on amorphous HEB$_2$, 5 folds by snapshot. Held-out RMSE a)~with all training frames, b)~with 2 training frames and c)~zero-shot. MP-0 small has no $l=1$ channels and therefore no bond projections.}
\label{fig:S1}
\end{figure*}

Fig.~\ref{fig:S1} compares eight foundation models, MACE-MP-0 small ($l_{max} = 0$), medium-0b3 and large\cite{macemp0}, MPA-0 medium\cite{macemp0,salex}, OMAT-0 medium\cite{macemp0,omat24}, MatPES-PBE-0 and MatPES-r$^2$SCAN-0\cite{matpes} and MACE-MH-1\cite{macemh1} in its single-head OMAT variant, each with the same pair model, Naik pretraining and calibrated fine-tuning, on the 508 development frames of amorphous HEB$_2$ (five folds by snapshot). The full-data values of this survey are for the pretrained protocol; the scratch values of the final model are in Table \ref{M-tab:final}. MP-0 large has $256\times0e + 256\times1o$ after its first layer, so it carries $l=1$ but no $l=2$ features, like all the others except MP-0 small, which has none. Unfreezing the last interaction and product block of OMAT-0 (2.08 M parameters, learning rate $10^{-4}$, 25 epochs) changed the full-data error from 0.0725 to 0.0711 eV against a frozen control run through the same code, at about 20 times the training cost.

\section*{Note S3. Maps of the descriptor space}

The generalized coordination number of a Pt atom in UMAP Figure (e) is the sum over its Pt neighbours within 3.35~\AA{} of their own Pt coordination numbers divided by 12; the height in UMAP Figure (f) is that of the bond midpoint above the mean height of the sheet atoms.

\section*{Note S4. Pt/Pd particles: DFT, MACE potential and coverage scan}

\begin{figure*}[!htbp]
\centering
\includegraphics[width=\textwidth]{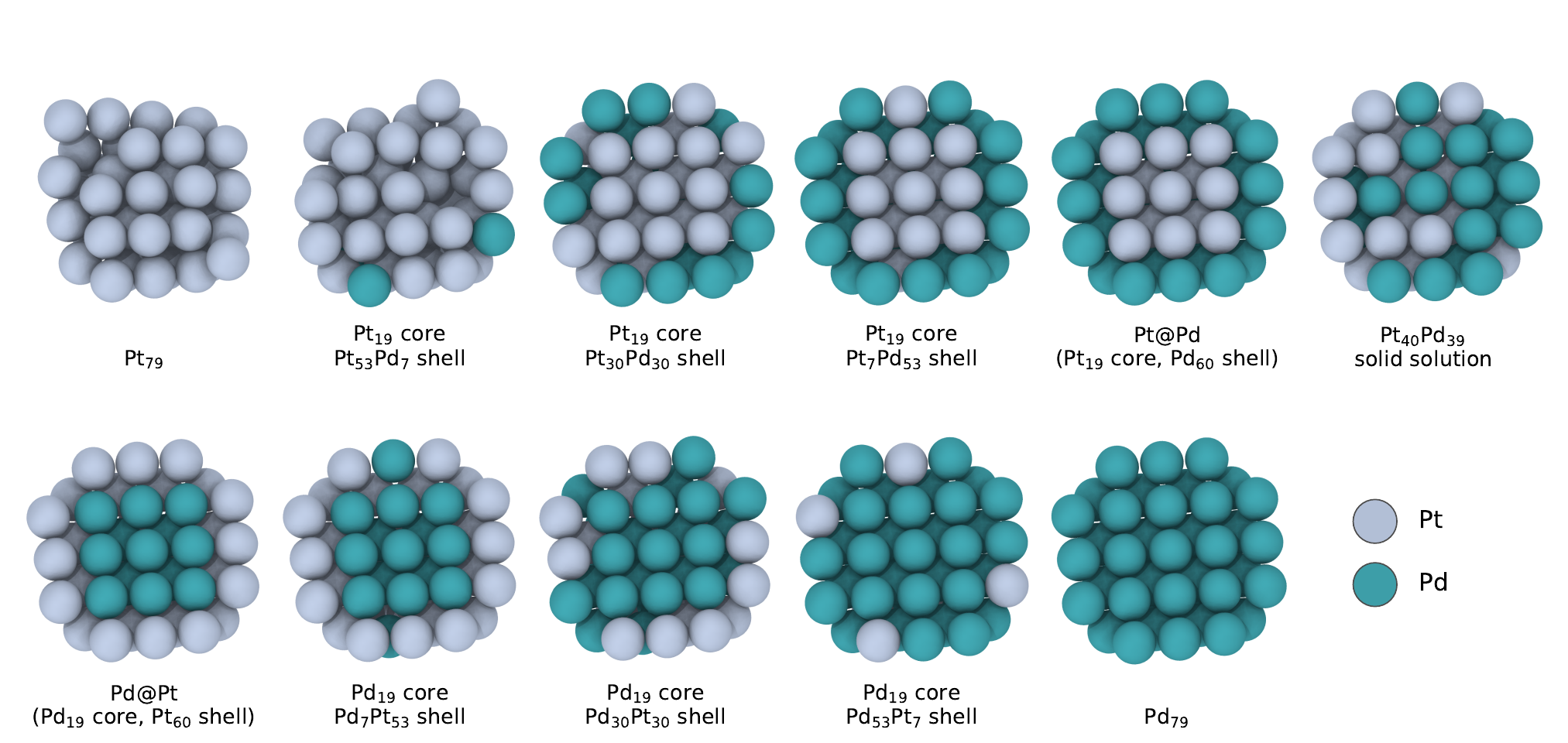}
\caption{The eleven Pt/Pd$_{79}$ particles, DFT-relaxed with 2 O$_2$ (O not shown), cut through the central atom by a \{100\} plane and seen on the cut face. Top row: Pt core under a shell from all Pt to all Pd, and the solid solution; bottom row: Pd core under a shell from all Pt to all Pd.}
\label{fig:Sptpdtypes}
\end{figure*}

\textit{DFT.} The relaxations used PREC = Normal, Methfessel--Paxton smearing of 0.05 eV, EDIFF = $10^{-4}$ eV and LREAL = Auto; the single points for the projection ISMEAR = 0 and PREC = Accurate. In the relaxed DFT structures at 30 O$_2$ the closest molecules are 2.75--2.93~\AA{} apart. The site assignment is a one-to-one matching of the centered metal positions to the reference lattice (the relaxed pure Pd particle at 2 O$_2$); a scan structure keeps the atom order of its DFT particle, so its sites are those of the DFT assignment and are not matched again.

\textit{MACE potential.} The fine-tuning set is every ionic step of the 33 relaxations with O$_2$ and of 112 relaxations of bare Pt/Pd$_{79}$ particles with the same settings, 5654 frames. Three relaxations with O$_2$ (one per coverage: 2 O$_2$ on the Pd core with a Pd$_{30}$Pt$_{30}$ shell, 15 O$_2$ on the Pt core with a Pt$_{30}$Pd$_{30}$ shell, 30 O$_2$ on the solid solution) and every eighth bare one were held out as the test set (552 frames); three others and another eighth of the bare ones were used for early stopping. A plain fine-tune of MPA-0 lost the short-range repulsion of the foundation model: relaxations with 30 O$_2$ collapsed (two atoms closer than 1~\AA), and a gentler plain fine-tune still failed on 6\% of the scan structures. The multi-head fine-tune replays 10\,000 Materials Project configurations through the foundation head (40 epochs, learning rate $10^{-4}$, EMA 0.99, energy and force weights 1 and 100), followed by 15 epochs with the 15- and 30-O$_2$ frames weighted 3 and 5. On the test set it gives 1.2--2.2 meV/atom and 36--63 meV/\AA{} for the three coverages and 1.4 meV/atom and 18 meV/\AA{} for the bare particles. The three held-out relaxations, started from their DFT starting structures, all converge, with mean M--M lengths within 0.006~\AA, M--O within 0.011~\AA{} and O--O within 0.002~\AA{} of DFT; the stock MPA-0 makes the M--O bonds 0.05--0.08~\AA{} too long.

\textit{Scan.} The metal frame of each particle was taken from its 2-O$_2$ starting structure; the molecules were put on random sets of bridge sites (pairs of neighbouring surface metals) with no metal shared, no two molecules closer than 2.9~\AA, and the mean DFT bridge geometry (M--O 1.97 and 2.01~\AA{} for Pt and Pd, O--O 1.387~\AA). The structures were relaxed with LBFGS to 0.05 eV/\AA{} (at most 1500 steps, step cap 0.1~\AA, stopped if two atoms come closer than 1~\AA): all 671 converged, in 91 steps and 9.6 s in the median, 2.4 GPU-hours in total on RTX PRO 4500 cards. Every particle was predicted by the fold model in which it was held out, for all pairs within 8~\AA, so every lattice bond of every structure is counted. 0.7\% of the lattice bonds are longer than 4.2~\AA, and their predictions, an extrapolation of the baseline, average $-0.22$ eV against $-0.17$ eV for DFT metal pairs at 4.0--4.2~\AA; setting them to zero changes the shell Pt--Pt change at 30 O$_2$ by 0.001 eV. Table \ref{tab:Sscan} compares the scan with DFT class by class, together with the held-out model on the DFT geometries. The MACE-relaxed particles have more stretched lattice bonds than DFT: 15--24 lattice bonds longer than 3.4~\AA{} per structure from 6 O$_2$ on, against at most 6 in DFT outside two particles that are distorted already at 2 O$_2$. Counted with a 3.4~\AA{} cutoff instead of the lattice, the shell Pt--Pt weakening at 30 O$_2$ would be 0.12 eV in the scan against 0.27 eV for the lattice count, and 0.20 against 0.22 eV in DFT.

\begin{table*}[!htbp]
\centering
\caption{Change of the class-mean ICOHP (eV) from 2 O$_2$ for the eight Pt/Pd particles with DFT at all three coverages: DFT labels; the held-out model (each particle predicted by the fold model that did not see it) on the DFT geometries; and the coverage scan with the same held-out models on the MACE geometries, mean of four random placements (scan at 15 O$_2$: mean of 14 and 16 O$_2$). Lattice bonds; M--O bonds within 2.4~\AA.}
\label{tab:Sscan}
\begin{hebtab}
\begin{tabular}{@{}|l|r|r|r|r|r|r|@{}}
\hebruleouter{1-7}
\textbf{Class} & \hebhead{DFT, 15} & \hebhead{model on DFT, 15} & \hebhead{scan, 15} & \hebhead{DFT, 30} & \hebhead{model on DFT, 30} & \hebhead{scan, 30} \\
\hebrulebold{1-7}
all M--M, core--core & $-0.011$ & $-0.004$ & $+0.022$ & $-0.015$ & $-0.006$ & $-0.004$ \\
\cline{1-7}
all M--M, core--shell & $-0.055$ & $-0.052$ & $-0.014$ & $-0.115$ & $-0.112$ & $-0.088$ \\
\cline{1-7}
all M--M, shell--shell & $+0.066$ & $+0.069$ & $+0.092$ & $+0.137$ & $+0.141$ & $+0.174$ \\
\cline{1-7}
Pt--Pd, core--core & $-0.027$ & $-0.024$ & $+0.018$ & $-0.015$ & $+0.002$ & $-0.004$ \\
\cline{1-7}
Pt--Pd, core--shell & $-0.037$ & $-0.033$ & $-0.024$ & $-0.111$ & $-0.106$ & $-0.090$ \\
\cline{1-7}
Pt--Pd, shell--shell & $+0.091$ & $+0.085$ & $+0.082$ & $+0.107$ & $+0.108$ & $+0.144$ \\
\cline{1-7}
Pt--Pt, core--core & $-0.002$ & $+0.012$ & $+0.068$ & $+0.024$ & $+0.034$ & $+0.035$ \\
\cline{1-7}
Pt--Pt, core--shell & $-0.142$ & $-0.135$ & $+0.006$ & $-0.151$ & $-0.152$ & $-0.078$ \\
\cline{1-7}
Pt--Pt, shell--shell & $+0.062$ & $+0.069$ & $+0.127$ & $+0.217$ & $+0.223$ & $+0.266$ \\
\cline{1-7}
Pd--Pd, core--core & $-0.010$ & $-0.006$ & $+0.003$ & $-0.027$ & $-0.019$ & $-0.018$ \\
\cline{1-7}
Pd--Pd, core--shell & $-0.021$ & $-0.025$ & $-0.019$ & $-0.112$ & $-0.113$ & $-0.086$ \\
\cline{1-7}
Pd--Pd, shell--shell & $+0.033$ & $+0.031$ & $+0.086$ & $+0.075$ & $+0.078$ & $+0.086$ \\
\cline{1-7}
M--O & $+0.287$ & $+0.294$ & $+0.131$ & $+0.403$ & $+0.416$ & $+0.294$ \\
\hebruleouter{1-7}
\end{tabular}
\end{hebtab}
\end{table*}

\section*{Note S5. Pretraining protocols with the distance baseline}

\begin{table*}[!htbp]
\centering
\caption{Held-out RMSE (eV) with MPA-0 scalar descriptors and the distance baseline, mean $\pm$ standard deviation over folds (number of folds in brackets; 10 for scratch, where a second random draw of the training structures was added), folds by structure: trained from scratch, pretrained on MattKeyBond, pretrained on both databases with the output calibration, and zero-shot from the latter. For HEB$_2$ the smallest $n$ is 2 snapshots; "all" is the four training folds.}
\label{tab:Sprotocols}
\begin{hebtab}
\begin{tabular}{@{}|l|c|r|r|r|r|@{}}
\hebruleouter{1-6}
\textbf{Application} & $n$ & \hebhead{scratch} & \hebhead{MattKeyBond} & \hebhead{both, calibrated} & \hebhead{zero-shot} \\
\hebrulebold{1-6}
Pt$_n$/graphene, Pt--Pt & 1 & 0.122 $\pm$ 0.008 (5) & 0.119 $\pm$ 0.013 (5) & \textbf{0.116 $\pm$ 0.011} (5) & 1.91 \\
\cline{1-6}
 & all & \textbf{0.071 $\pm$ 0.014} (5) & 0.072 $\pm$ 0.014 (5) & \textbf{0.071 $\pm$ 0.014} (5) & \\
\hebrulebold{1-6}
Pt/Pd + O$_2$ (first set), M--M & 1 & 0.391 $\pm$ 0.092 (10) & 0.244 $\pm$ 0.024 (5) & \textbf{0.237 $\pm$ 0.006} (5) & 1.29 \\
\cline{1-6}
 & all & 0.054 $\pm$ 0.024 (10) & 0.048 $\pm$ 0.015 (5) & \textbf{0.045 $\pm$ 0.015} (5) & \\
\hebrulebold{1-6}
amorphous HEB$_2$, all pairs & 2 & 0.513 $\pm$ 0.055 (10) & 0.467 $\pm$ 0.045 (5) & \textbf{0.456 $\pm$ 0.042} (5) & 1.15 \\
\cline{1-6}
 & all & \textbf{0.154 $\pm$ 0.002} (10) & 0.157 $\pm$ 0.002 (5) & 0.156 $\pm$ 0.002 (5) & \\
\hebruleouter{1-6}
\end{tabular}
\end{hebtab}
\end{table*}

In joint training a 16-entry one-hot of the source was appended to $\mathbf{x}_{ij}$ (it is zero in all other runs). Table \ref{tab:Sprotocols} is the first protocol study, with MPA-0 scalar descriptors and the distance baseline, run before the final model and before the second Pt/Pd set existed. Its Pt/Pd row is the first Pt/Pd set: the same eleven particles with one O$_2$ each, projected with LOBSTER and, since its input listed no basis functions, a minimal $spdf$ basis with a charge spilling of 40--49\%, so the errors there measure how well the model reproduces these numbers and not the bonds. Its Pt/graphene folds are by structure, not by cluster. Zero-shot with the distance baseline uses the polynomial of the database for the same pair type where it exists and the mean database label where it does not. Pretraining paid only with one or two training structures; with both databases and the output calibration it was never worse than with MattKeyBond alone, although the differences are within the fold spread. Joint training on the databases and the application in one model was worse than training from scratch at every $n$, since the application pairs are about $10^{-4}$ of the loss.

\section*{Note S6. Timing}

LOBSTER and VASP times are the wall times printed in the \texttt{lobsterout} and OUTCAR files of the 24 ordered cells, run on 2$\times$Xeon Gold 6248 with 12--20 threads for LOBSTER and 12--20 MPI ranks for VASP. The machine ran two such jobs at a time, so these times are on the high side. gpucohp times are the wall times in its \texttt{lobsterout} files, from reading the wavefunction to writing all outputs, for one $\Sigma 8$, one $\Sigma 7$ and one $\Sigma 6$ five-metal cell on one RTX 2080 Ti ($\Sigma 7$ on a GTX 1080 Ti) and for the 1246 amorphous frames on the same two GPU types. The model was timed from atoms to ICOHP of all pairs within 4.2~\AA{} (descriptors, neighbor list, three heads), median of three repeats, on six CPU threads of a 12-core workstation and on one RTX 4090, for the same 24 cells, 100 amorphous frames and supercells of the $\Sigma 7$ HEB$_2$ cell of up to 3744 atoms. A 3744-atom cell was not attempted with LOBSTER. Per boundary cell the model on six CPU threads is $2.8\cdot10^4$ ($\Sigma 8$) to $1.4\cdot10^5$ ($\Sigma 6$) times faster than LOBSTER, and gpucohp is 230--360 times faster than LOBSTER (243 times for the $\Sigma 8$ cell projected by both codes from the same wavefunction).

\section*{Note S7. DFT and projection settings}

\begin{table*}[!htbp]
\centering
\caption{DFT and projection settings of the label sets. All projections use the pbeVaspFit2015 basis; spins are summed where ISPIN = 2.}
\label{tab:Ssettings}
\begin{hebtab}
\begin{tabular}{@{}|l|l|l|@{}}
\hebruleouter{1-3}
\textbf{Set} & \textbf{DFT} & \textbf{Projection and checks} \\
\hebrulebold{1-3}
amorphous HEB$_2$ & VASP 6.4.3, PBE, ENCUT 500 eV, & gpucohp; 0.0006 eV RMSE against \\
 & Gaussian smearing 0.05 eV, ISYM $=-1$, & LOBSTER 5.1.1 on a 48-atom boundary cell; \\
 & NBANDS = size of the local basis, & the coarse mesh changes ICOHP by \\
 & KSPACING 0.30~\AA$^{-1}$ & 0.004 eV on average at an eighth of the cost \\
\cline{1-3}
ordered cells & VASP, PBE, fully relaxed; & LOBSTER 5.1.1 \\
 & single points with KSPACING 0.16~\AA$^{-1}$ & \\
\cline{1-3}
Pt$_n$/graphene & VASP, PBE, ENCUT 300 eV, & LOBSTER \\
 & no spin polarization & \\
\cline{1-3}
Pt/Pd$_{79}$ + $n$O$_2$ & VASP, RPBE-D3 (GGA = RP, IVDW = 11), & gpucohp; charge spilling 0.60--1.04\% \\
 & ISPIN = 2, ENCUT 480 eV, $\Gamma$ point, & \\
 & 21~\AA{} box (further settings in Note S4) & \\
\hebruleouter{1-3}
\end{tabular}
\end{hebtab}
\end{table*}

\bibliography{refs}